\documentclass[aps,twocolumn,showpacs,floats,superscriptaddress]{revtex4}
\usepackage{graphicx}
\usepackage{amssymb,amsmath}
 \topmargin=-5pt

\begin{document}

\title{Monte Carlo study of two-dimensional Bose-Hubbard model}

\author{Barbara Capogrosso-Sansone}
 \affiliation{Department of Physics, University of
Massachusetts, Amherst, MA 01003}

\author{\c{S}ebnem G\"{u}ne\c{s} S\"{o}yler}
 \affiliation{Department of Physics, University of
Massachusetts, Amherst, MA 01003}
\author{Nikolay Prokof'ev}
\affiliation{Department of Physics, University of Massachusetts,
Amherst, MA 01003}
 \affiliation{Russian Research Center
``Kurchatov Institute'', 123182 Moscow, Russia}

\author{Boris Svistunov}
\affiliation{Department of Physics, University of Massachusetts,
Amherst, MA 01003} \affiliation{Russian Research Center
``Kurchatov Institute'', 123182 Moscow, Russia}

\begin{abstract}
One of the most promising applications of ultracold gases in optical
lattices is the possibility to use them as quantum emulators of more
complex condensed matter systems. We provide benchmark calculations,
based on exact quantum Monte Carlo simulations, for the emulator to
be tested against. We report results for the ground state phase
diagram of the two-dimensional Bose-Hubbard model at unity filling
factor. We precisely trace out the critical behavior of the system
and resolve the region of small insulating gaps, $\Delta\ll J$. The
critical point is found to be $(J/U)_c=0.05974(3)$, in perfect
agreement with the high-order strong-coupling expansion method of
Ref.~\cite{Monien}. In addition, we present data for the effective
mass of particle and hole excitations inside the insulating phase
and obtain the critical temperature for the superfluid-normal
transition at unity filling factor.
\end{abstract}
\parindent 8.mm
\maketitle

\begin{figure}[t]
\centerline{\includegraphics[angle=0,scale=0.49,height=2.5in]
{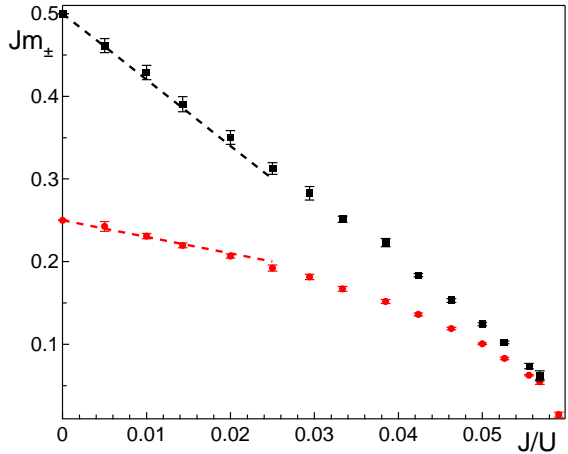}} \caption{(Color online) Effective mass for
particle (circles) and hole (squares) excitations as a function of
$J/U$. The exact results at $J/U=0$ are $m_{+} =0.25/J$ and $m_{-}
=0.5/J$. By dashed lines we show the lowest order in $J/U$
correction to the effective masses. Close to the critical point
the two curves overlap, directly demonstrating the emergence of
the particle-hole symmetry. At $J/U=0.059$, the sound velocity is
$c/J=4.8\pm 0.2$. } \label{eff_mass}
\end{figure}

In the last few years, manipulation of quantum gases in optical
lattices has been characterized by fast and striking advances in
trapping techniques (see e.g. \cite{Bloch} \cite{Ketterle} and
\cite{review} for a review), with the main remaining challenge
being the addressability of single sites. It is towards this
direction that some experimental groups are devoting their efforts
nowadays \cite{GreinerWeb}. Access to a single site would enable
\emph{in situ} measurements of observables of interest and direct
measurement of correlation functions, the knowledge of which would
ease the study and characterization of new exotic states of
matter. While ultracold gases in optical lattices are of interest
on their own, one could also think of a broader and more ambitious
project of using such systems as quantum simulators of
difficult-to-solve condensed matter systems and models. One
prominent example is quantum magnetism in electronic systems which
may be relevant to high $T_c$ superconductivity. Since such
systems are theoretically hard to address, one could alternatively
think of mimicking models of interest (Hubbard models, for
example) with ultracold gases in optical lattices.\\ \indent It is
in this framework that the DARPA agency has developed and funded a
program whose goal is building fermionic and bosonic optical
lattice emulators. There are two main challenges to meet:
addressability of single sites and engineering of exchange
interaction among atoms. The addressability of single lattice site
is crucial not only for local measurements but also for
manipulation of single atoms, which would open up the way to
applications in quantum computing. Engineering spin exchange
interactions is essential in order to study quantum magnetism. It
has already been shown that two-component boson systems with
properly tailored exchange interactions, can be used to realize
quantum spin Hamiltonians \cite{Kuklov,Demler1,Demler2}.
Altogether, the optical lattice emulator, the first example of the
special purpose quantum simulator, would enable one to explore new
exotic states and answer open questions in the fields of quantum
magnetism and superconductivity, including the interplay between
the two (e.g.
by determining ground states of Hamiltonians with competing orders).\\
\indent Within the quantum gas microscope implementation
\cite{GreinerWeb}, individual atoms are magnetically transported
from a 2D surface trap in the focal plane of an ultra-high aperture
objective to a spatially separated vacuum chamber \cite{Greiner}.
The most natural first step for understanding advantages and
limitations of this technique of atom imaging is to calibrate it
against the simplest correlated 2D system, the Bose-Hubbard
Hamiltonian on the square lattice:
\begin{equation}
H\; =\; -J\sum_{<ij>} b^{\dag}_i\,b_j +\frac{U}{2}\sum_i
n_i(n_i-1) -\sum_i \mu_i n_i\; , \label{BH}
\end{equation}
where $b^{\dag}_i$  and $b_i$ are the bosonic creation and
annihilation operators on the site $i$, $ J$ is the hopping matrix
element, $U$ is the on-site repulsion and $\mu_i=\mu-V(i)$ is the
difference between the global chemical potential $\mu$ and the
confining potential $V(i)$. At zero temperature and integer filling
factor, the system features the superfluid(SF)-to-Mott-insulator(MI)
phase transition \cite{Fisher}, with the MI phase being uniquely
characterized by the energy gap $\Delta$ to create a particle-hole
excitation. The ground state phase diagram of the homogeneous system
(in the $\mu/U$ vs $J/U$ plane) has a characteristic lobe shape with
the
system being in the MI state inside the lobe and SF outside. \\
\indent In experiments, gases in optical lattices are confined by an
external potential. So far, this has resulted in limitations in the
observation of a quantum phase transition due to measurement
averaging over the whole cloud. With the high-resolution quantum gas
microscope, measurements can be performed locally and averaging over
the inhomogeneous system can be avoided. The first goal of the
bosonic quantum emulator is to map out the ground state phase
diagram. The standard approach is based on a local chemical
potential approximation where the density of the homogeneous system
with the chemical potential
\begin{equation}
\mu_{i}^{\rm (eff)}\; =\; \mu \, - \, V(i)\; , \label{effective
mu}
\end{equation}
is identified with the density at the site $i$ of the
inhomogeneous system.\\
\indent In this paper we provide benchmarks for the bosonic
quantum emulator to be tested against. We report results of
large-scale exact quantum Monte Carlo simulations of model
\ref{BH}, by the Worm algorithm \cite{worm}. We focus on the
homogeneous case and unity filling factor. Worm algorithm allows
efficient sampling of the single-particle Green function. Precise
data for the Green function enable us to carefully trace out the
critical behavior of the system and resolve the phase diagram in
the region of small insulating gaps, $\Delta\ll J$. We also
present data for the effective mass of particle and hole
excitations inside the insulating phase. Effective masses
characterize the phase transition away from the tip of the lobe.
Here the transition is described by the physics of the weakly
interacting Bose gas in the limit of vanishing density
\cite{Fisher}.\\ \indent In order to completely characterize the
system the full phase diagram in the parameter space
$(\mu/U,J/U,T/J)$, where $T$ is the temperature, is needed. Here
we limit ourselves to studying ground state properties and
calculating the critical temperature for the SF-normal transition
at unity filling factor. An exhaustive finite temperature study of
the system is in
progress in another group \cite{Pollet_in_progress}. \\
\begin{figure}[t]
\centerline{\includegraphics[angle=0,scale=0.55]
{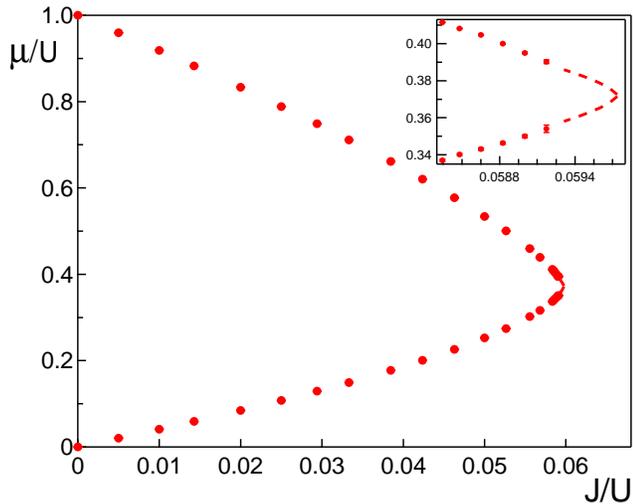}} \caption{(Color online) Phase diagram of the
first MI-SF lobe. Solid circles are numerical data, with error bars
shown but barely visible. The inset is a blow up of the region close
to the tip. Dashed lines represent the critical region as calculated
from finite size scaling. } \label{phase_diagram}
\end{figure}
\begin{figure}[t]
\centerline{\includegraphics[angle=0,scale=0.45]
{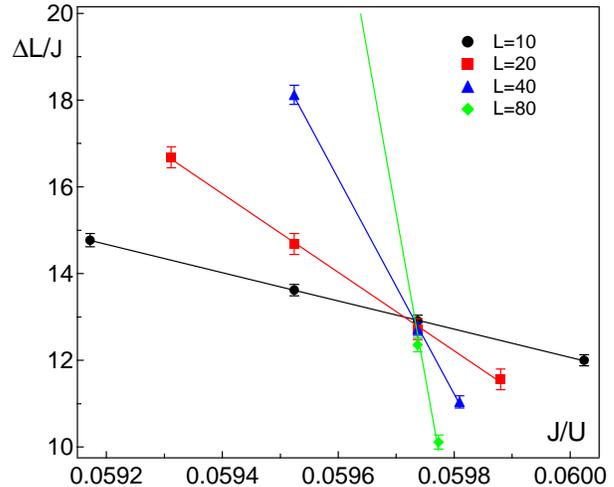}} \caption{(Color online) Finite size scaling
of the energy gap at the tip of the lobe. Lines represent linear
fits used to extract the critical point. The critical point can be
directly read from the intersection of the curves:
$(J/U)_c=0.05974(3)$.} \label{critical_point}
\end{figure}
\indent We now turn to the presentation of our results. The
procedure used to determine the ground state phase diagram and
extract effective masses of particle and hole excitations from the
Green function was discussed in details in Ref.~\cite{BH_3D}. In
Fig.~\ref{phase_diagram} we present results for the ground state
phase diagram corresponding to unity filling. The inset shows the
region around the tip. Circles represent the simulation data while
dash lines are obtained from the finite size scaling analysis.
Simulations were done for linear system sizes $L=10,\;20,\; 40,\;
80$. We do not see any significant size effect up to $J/U\sim
0.057$. In order to extract the position of the critical point at
the tip of the lobe and determine the extension of the critical
region, the standard finite size scaling argument was used (see
Ref.~\cite{BH_3D}), with the critical exponent for the correlation
length $\nu=0.6715$. The finite size scaling of the energy gap is
presented in Fig.~\ref{critical_point}. One can directly read the
position of the critical point from the intersection of the curves:
\begin{equation}
(J/U)_c=0.05974(3)\;\;\;\;(n=1).\label{critical_point_eq}
\end{equation}
\\ \indent Equation (\ref{critical_point_eq}) and Fig.~\ref{phase_diagram}
constitute the most precise quantum Monte Carlo simulation for the
Hamiltonian \ref{BH} which is in perfect agreement with the result
of Ref.~\cite{Monien}, where the authors carried out a strong
coupling expansion up to 13-th order. Note that the critical region
in Fig.~\ref{phase_diagram} is resolved with accuracy $\ll J$, i.e.
for gaps  $\Delta<J$, which is crucial for studies of the emerging
relativistic physics at the lobe tip.
\\ \indent In Fig.~\ref{eff_mass} we plot effective masses for particle (circles)
and hole (squares) excitations. Dispersion relations were fitted by
a parabola, with the exception for $J/U=0.059$ where we used a
relativistic dispersion relation. Close to the tip of the diagram,
the action is isotropic in space and imaginary time, giving rise to
a relativistic behavior \cite{Fisher}. In the limit
$J/U\rightarrow0$, where one can calculate effective masses
perturbatively, our data converge to the analytical result (dashed
lines). To the first order, the perturbative expansions are given by
(we set the lattice pariod and Planck's constant equal to unity):
\begin{equation}
Jm_+=0.25-2J/U,\;\;\;\;\;Jm_-=0.5-8J/U\label{eff_mass_eq}
\end{equation}
for particle and hole excitations respectively. On approach to the
critical point, instead, data for particle and hole excitations are
merging together, in agreement with the emergent particle-hole
symmetry in the critical region. From the fit done at $J/U=0.059$
using the relativistic dispersion relation, we extract the value of
the sound velocity $c/J=4.8\pm 0.2$ and effective mass
$m_*=0.015\pm0.0015$. We would like to point out that effective
masses can also be extracted using the method of Ref.~\cite
{Monien}, although we did not find any calculation in the
literature.
\begin{figure}[t]
\centerline{\includegraphics[angle=-90,width=3.5in]{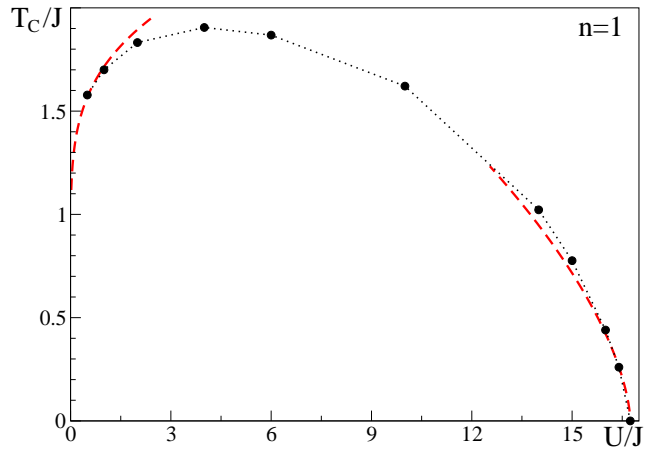}}
\caption{(Color online) Finite-temperature phase diagram at
filling factor $n=1$. Solid circles are simulation results (the
dotted line is to guide an eye), error bars are plotted. Dashed
lines are analytical results for the weakly interacting gas [see
Eq.~(\ref{T_c_lowU})] and for the strongly interacting gas close
$(J/U)_c$. } \label{tc}
\end{figure}
\\ \indent In Fig.~\ref{tc} we show the phase diagram for the
SF-normal transition at integer filling factor $n=1$. The transition
is of Berezinskii-Kosterlitz-Thouless type \cite{KT}. The critical
temperatures were found from extrapolation to infinite system size
of the standard finite size scaling for the Kosterlitz-Thouless
transition (see e.g. Ref.~\cite{KT_scaling}). In the figure, circles
are numerical results and dashed lines are analytical expressions in
the two limiting cases. In the weakly interacting regime the
critical temperature is given by:
\begin{equation}
T_c=\frac{2\pi n}{m\ln(\xi/mU)},\label{T_c_lowU}
\end{equation}
where $n$ is the density ($n=1$ in this case), $m$ is the mass
($m=1/2J$ in the lattice), and $\xi$ is a dimensionless parameter
which was found numerically in Ref.~\cite{KT_scaling} to be
$\xi=380\pm 3$. On the approach of the critical point, instead,
one can use the following scaling argument. Close to the critical
point the superfluid density is $\rho_s(T=0)\sim \xi^{-1}\sim
t^\nu$, where $t= (U/J-(U/J)_c)$. Under the assumption
$\rho_s(T=0)\sim\rho(T_c)$, which should hold for low enough
critical temperatures, i.e. close to the critical point $(U/J)_c$,
one concludes that $T_c\sim\rho(T_c)\sim t^\nu$. The dashed line
in the plot is a fit done using the function $f(x)=At^\nu$, where
A=0.49(2) is a fitting parameter. On both sides, numerical results
clearly converge to the analytical expressions.
\\ \indent Many interesting phenomena
happening at zero or nearly zero temperature have not been
observed yet. This is because, so far, it has been a challenge for
experimentalists to reach low enough temperatures. In order to
overcome this challenge, one can exploit the inhomogeneity of the
entropy distribution of the harmonically and optically trapped
gas. The idea has been originally proposed in
Ref.~\cite{cooling_Cirac1} and \cite{cooling_Cirac2}, where the
authors suggest several cooling protocols. Some of the protocols
make use of the filtering scheme of Ref.~\cite{filtering}, others
require spin dependent lattices. In all the protocols the main
idea is to  \emph{relocate} the entropy by removing a small
fraction of the particles carrying almost all the entropy.
Recently it has also been suggested a cooling protocol based on
coupling entropic particles with a system at lower temperature
(i.e. a ``refridgerator'') \cite{cooling_Ho}. Having in mind a
setup, similar to the one described in Ref.~\cite{cooling_Cirac1}
and \cite{cooling_Cirac2}, we would like to suggest a simple and
efficient cooling protocol which does not require coupling to a
refridgerator or exciting particles to a different internal energy
level. Consider a system with $U\gg |\mu_{i_0}-\mu_{i=0}| ,J, T$
and $|\mu_i-\mu_{i+1}|\gg J$, where $\mu_i$ is given by
Eq.~(\ref{effective mu}), $i=0$ is at the potential minimum while
$i \sim i_0$ corresponds to the boundary of the system. The
condition $U\gg |\mu_{i_0}-\mu_{i=0}|, J$ means that the
groundstate density is essentially uniform. The condition $U \gg
T$ guaranties that the entropy is located in a thin peripheral
region of the system. In view of the condition
$|\mu_i-\mu_{i+1}|\gg J$, the elementary excitations of the system
are single-site (localized) particles and holes obeying Fermi
statistics (in the real space)~\cite{cooling_Cirac1}.
\begin{figure}[t]
\hspace*{+0.9cm}
\centerline{\includegraphics[height=2.5in,width=3.5in]{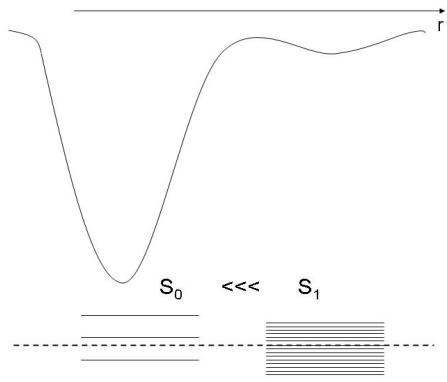}}
\caption{Sketch of the proposed cooling protocol. On top, a shallow
and a deep trap are superimposed. At the bottom, the corresponding
densities of states are sketched. The dashed line represents the
Fermi level. Adiabatically displacing the shallow trap results in
displacing entropic particles with consequent dramatic decrease of
entropy per particle inside the deep trap.} \label{density_states}
\end{figure}
\\ \indent Take two superimposed traps, a very steep one and a very
shallow one. The density of energy levels of the shallow trap is
very high as compared to that of the steep one. The shallow trap
thus carries most of the entropy. Adiabatically displacing the
shallow trap would then result in displacing ``entropic" particles
and ultimately separating them from the particles of the steep trap.
Academically speaking, the very last stage of this process cannot be
adiabatic in view of the exponentially suppressed tunnelling between
two traps. However, this non-adiabaticity simply means that the two
systems become independent, so that the (now dramatically decreased)
entropy of the steep trap has nothing to do any longer with the
state of the particles in the shallow trap. The procedure is
sketched in Fig.~\ref{density_states}. For better results, one can
perform a number of cooling cycles consisting of the above-described
entropy relocation procedure followed by adiabatically increasing
(and subsequent decreasing) the $J/U$ ratio in order to lift the
localization constraint preventing redistribution of tiny fraction
of remanent particle and hole excitations in the bulk towards the
perimeter. Another possibility for entropy relocation consists of
progressively lowering the confinement below the Fermi level in
order to evaporate the entropic particles from the system.
\\ \indent In conclusion, we have presented numerical results for the single
component Bose-Hubbard model. We have confirmed previous
calculations \cite{Monien} for the critical point at the tip of the
$n=1$ lobe and presented results for the effective masses for
particle and hole excitations. We have determined the critical
temperature for the SF-normal transition for the unity filling case.
These benchmark calculations provide a robust test for the
bosonic optical lattice emulator.\\
\indent We are grateful to Tin-Lun Ho for useful discussions. This
work has been supported by the DARPA OLE program.

\end{document}